\documentclass[a4paper]{article}
\usepackage{multirow}
\usepackage{INTERSPEECH2021}
\usepackage{cite}
\usepackage{indentfirst}
\usepackage{lettrine}
\usepackage{url}
\title{Complex Spectral Mapping With  Attention Based Convolution Recurrent Neural Network for Speech Enhancement}
\name{Liming Zhou$^1$, Yongyu Gao$^1$,Ziluo Wang$^1$,Jiwei Li$^1$,Wenbin Zhang$^1$}
\address{
  $^1$CloudWalk Technology Co., Ltd., Shanghai, China
  }
\email{$^1$\{zhouliming,gaoyongyu,wangziluo,lijiwei,zhangwenbin\}@cloudwalk.cn}

\begin{document}

\maketitle
\begin{abstract}
  Speech enhancement has benefited from the success of deep learning in terms of 
  intelligibility and perceptual quality. Conventional time-frequency (TF) domain
  methods focus on predicting TF-masks or speech spectrum,via a naive convolution
  neural network or recurrent neural network.Some recent studies were based on 
  Complex spectral Mapping convolution
  recurrent neural network (CRN) .
  These models skiped directly from  encoder layers' output and decoder
  layers' input ,which maybe thoughtless. We proposed an attention mechanism based skip connection
  between encoder and decoder layers,namely Complex Spectral Mapping With Attention
  Based Convolution Recurrent Neural Network (CARN).Compared with CRN model,the proposed CARN model improved more than 10\% relatively at several metrics such as PESQ,CBAK,COVL,CSIG and son,and 
  outperformed the place 1st model in both real time and non-real time track of the DNS Challenge 2020 at these metrics.
  
  
\end{abstract}

\section{Introduction}
\lettrine[lines=2]{S}{peech} enhancement technology is essential to improve the intelligibility and quality of noisy speech signal\cite{speech_enhancement}.
Classcial speech enhancments techniques include spectral subtraction\cite{1163209}, Wiener filtering\cite{1455809}, 
miniimum mean-square- error(MMSE) estimator\cite{1164453} and the optimally modified log-spectral amplitude speech estimator\cite{2001Speech}.
These convetional methods based on time-frequency domain achieve relatively good performance in stationary-noise environment,
whereas they are not robust enough in tackling non-stationary noises in most scenes.

Over past few years, deep nerual networks (DNNs) have significantly evaluated the performance of speech enhancement\cite{2017Supervised}. 
Existing DNN apporaches provide better results compared to classic techniques. In \cite{DBLP:conf/interspeech/MaasLOVNN12}\cite{2017A}, 
recurrent nerual network (RNNs) is advanced to model temporal features. Auto-encoder is employed for speech enhancement in  \cite{DBLP:conf/interspeech/LuTMH13},
and the first speech enhancemen benchmark ultilizes DNN as non-linear regression function\cite{2015A}.
A convolutional recurrent neural  network (CRN) is used to extract long-context information\cite{inproceedings222}. 
Speech enhancement techniques also take example by image synthesis that using Generative adversarial network(GAN) 
architecture to reconstruct target speech signal \cite{2017SEGAN,2019MetricGAN,Lin2019,2019Sergan,DBLP:conf/interspeech/LinNWMSW20}.
These DNN supervised-based speech enhancements methods are outperform than classic algorithms.

Recently, U-net structures have achieved significant success and have overran their performance than basic DNN architectures
in various machine learning tasks including medical diagnostics\cite{2015U}, semantic segmentation\cite{2017SegNet}, 
singing source separation\cite{2017Sing} and others. Motivated by this sucess, speech enhancement has explorerd 
U-net structures in both raw waveform\cite{2018WaveUnet,2018ImprovedSE,2018SpeechDerevb,2017SEGAN} and time-frequency features\cite{2018TimeF}.
In \cite{A2019Demucs}, Wave-U-Net employed GLU activation in encoder and decoder, as well as bidirectional LSTM in-between.
DCCRN\cite{2020DCCRN}takes advantages of u-net and deep complex network \cite{DCN} for denoising. 
The combination between attention-unit and U-net boost the performance of speech enhancment one step further. 
Self-attention is an efficient context information aggregation mechanism that operates on the input sequence itself
and that can be utilizedfor any task that has a sequential input and output. Attention wave-u-net \cite{2019Attention} outperforms all other
published speech enhancement approaches on the Voice BankCorpus (VCTK) dataset.

Mask-based target, which describes the time-frequency relationships between clean speech and background noise,is utilized to train the network. Typically, conventinal masks consist of ideal binary mask(IBM)\cite{IBM}, ideal ratio mask(IRM)\cite{IRM} and spectral magnitude mask(SMM)\cite{SMM} only consider the magnitude between clean
waveform and mixture audio. Subsequently, phase is taken into  account, where phase-sensitive mask(PSM)\cite{PSM} is the first method to show the feasibility of phase information, and complex ratio mask (CRM)\cite{CRM} is announced 
that it can reconstruct speech perectly by enhancing both real and imaginary components of the division of clean speech and mixture speech spectrogram simultaneously. Soon afterwards, CRN\cite{CRN} used one 
encoder and two decoders for complex spectral mapping(CSM) to coincidently estimate the real and imaginary spectrogram of mixture speech.
It isworth noting that CRM and CSM possess the full informationof a speech signal 
so that they can achieve the best oracle speech enhancement performance in theory.

In this paper, we propose a network called CARN incorporating effitive components of u-net, 
attention mechanism, skip connection, LSTMs in between encoder-decoder and CRM in time-frequency domain. 
We also Investigated conbining gate convolution network with CARN model into a model,named GCARN.
We show that, based on the speech quality metrics (PESQ, etc.) on the 2020 DNS challenge\cite{reddy2020interspeech}, 
and the dataset released by Valentini\cite{valentini2017noisy}.
the attention-lstm-unet utilized crm in time-frequency domain achieve significant imporvement 
results outperforming others published speech enhancement methods on these datasets.


\section{The CARN Model}
The CRN model was firstly proposed by Ke Tan\cite{tanCRN},and was further investigated with complex spectral mapping\cite{CRN} mechanism and gate convolution\cite{GCRN}.
According to the above research, we investigated a new model named CARN,which was combined CRN with attention mechanism.In the CARN model.we used attention based skip connections from encoder layers to decoder layers.
Furthermore,we constrast our proposed model with the gate convolution based CRN model,as well as several  previous works.

\subsection{The CARN Architecture}
\begin{figure}[bthp]
  \centering
  \includegraphics[width=0.6\linewidth]{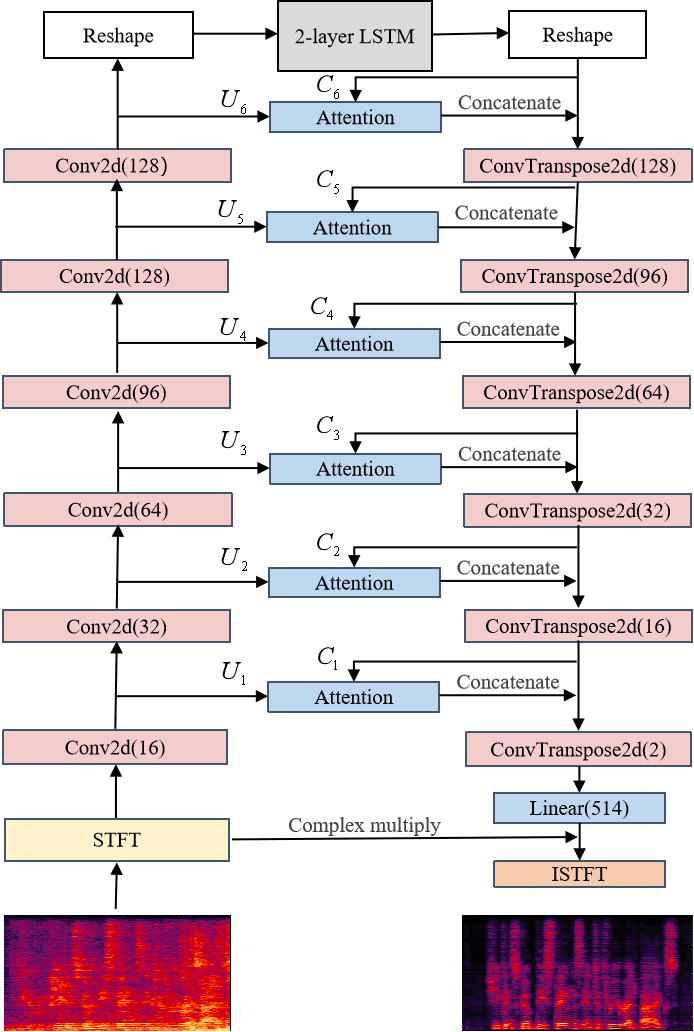}
  \caption{CARN network}
  \label{fig:carn_fig}
\end{figure}

The encoder and decoder both consist of 6 Conv2d blocks with PReLU activation function,aimming at extracting high-demensional features
from the input features as well as reducing the resolution.Between encoder and decoder,we used two LSTM layers to study temporal features. The model is illustrated as Figure~\ref{fig:carn_fig}.We took spectrogram feature as input.
The LSTM layers hidden size is 512,The T-F kernel size is 3, stride is $1*2$, 
for each Conv2d or ConvTranspose2d layer. Each Conv2d or ConvTranspose2d layer is followed by a batchnorm layer. A linear layer is embedded after last ConvTranspose2d layer to map the complex ratio mask(CRM) from the ouput features.
At last,CRM multiply with the input stft spectrogram to get clean stft spectrogram,refering to \eqref{4} and \eqref{5}.
All the activation function is PReLU.

\subsection{Attention Mechanism }

Different from conventional CRN architecture, self-attention masks are applied to multiply with the output of encoder by skip-connection. The output of attention layer is concatenated to the last decoder output for the next decoder input.

\begin{figure}[bthp]
  \centering
  \includegraphics[width=0.6\linewidth]{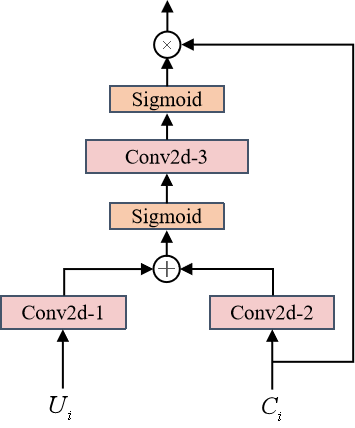}
  \caption{attention}
  \label{fig:attention_fig}
\end{figure}

The AttentionBlock\cite{2019Attention} is described in Figure~\ref{fig:attention_fig}. $ U_{i} $ is
the output of encoder architecture and $ C_{i} $ is the output of LSTM layers or decoder covolution layers. 
 Additional two  2-d convolutions, with kernel size 3, output channels twice of input channels,
 refering as $ W_{g} $ and $ W_{x} $, which are used to mapping $U_{i}$ and $C_{i}$ 
 to high-dimensional space feature,are used to model the attention mechanism. The high-dimensional space feature
 demension is  twice of the demension of $C_{i}$;
 The high-dimensional space feature layer output can be discriped as \eqref{1}.\par
\begin{equation}
        A_{i}=\sigma (W_{g}\otimes U_{i}+W_{x}\otimes C_{i})\label{1}
\end{equation}
where $ U_{i} $ and $ C_{i} $ present $ i th $ layer of encoder and decoder, respectively.
 $\sigma$ is sigmoid fuction. The output of self-attention block,
\begin{equation}
        B_{i}=\sigma (W_{f}\otimes B_{i})\cdot C_{i}\label{2}
\end{equation}\par
where $W_{f}$ presents another 2-d concolution layer.

\subsection{Training targets}
CARN estimates CRM and is optimized by signal approximation (SA).Given the complex-vauled STFT spectrogram of clean speech S and 
noisy speech Y,CRM can be defined as 

\begin{equation}
  CRM = \frac{Y_rS_r+Y_iS_i}{Y_r^2+Y_i^2}+j\frac{Y_rS_i-Y_iS_r}{Y_r^2+Y_i^2}\label{3}
\end{equation}
where \(Y_r\) and \(Y_i\) refer as the real and imaginary parts of the noisy complex 
spectrogram,respectively.\(S_r\) and \(S_i\) refer as the real and imaginary parts of the 
clean complex spectrogram, similarly.Let \(\hat{S}_r\) and \(\hat{S}_i\) denote 
the real and imaginary parts of the estimating denoise audio complex spectrogram,respectively.
\(\hat{M_r}\) and \(\hat{M_i}\) denote the real and imaginary parts of CRM,then
\begin{equation}
  \hat{S_r} = \hat{M_r}Y_r-\hat{M_i}Y_i \label{4}
\end{equation}
\begin{equation}
  \hat{S_i} = \hat{M_r}Y_i+\hat{M_i}Y_r     \label{5}
\end{equation}

\subsection{Loss function}
We train the model with loss function as 
\begin{equation}
  Loss(\hat{S},S) = (|\hat{S}|^{0.3}-|S|^{0.3})^2+0.2*|\hat{S}^{0.3}-S^{0.3}|^2
\end{equation}
where \(\hat{S}\) and \(S\) denote estimating denoise audio and clean audio respectively.  \(S^{0.3}=|S|^{0.3}e^{j\angle S}\) ,is the power-compressd STFTs.
This loss function consists of spectrogram MSE (mean square error) and power-compressd
STFTs MSE.

\section{Experiemnt}
\subsection{Datasets}
In our experiments, we evaluate the proposed models on two datasets.
\subsubsection{Dataset 1: Noisy speech database}
The first dataset\footnote{https://datashare.ed.ac.uk/handle/10283/2791}, is released by Valentini et. al.\cite{valentini2017noisy}, widely used by speech enhancement research, and it generalizes on various types of noise for different speakers. This dataset includes 
 clean and noisy audio data at 48kHz sampling frequency, which requires downsampling to 16kHz for training and testing. The clean sets are recording
 of sentences, sourced from various text passages, and thirty English-speakers, including male and female with various accents, are selected from the Voice Bank corpus\cite{veaux2013voice}. 28 and 2 speakers are assigned to the training and test sets, respectively.
 The test set consists of 20 different noise conditions, 5 type of noise sourced from the DEMAND database ,yielding 824 test items, with approximately 20 different sentences in each condition per test speaker.\cite{valentini2017noisy}
 \subsubsection{Dataset 2: DNS 2020}
 The second dataset is based on the data provided by
the Interspeech 2020 DNS Challenge dataset \cite{reddy2020interspeech}.
The DNS Challenge dataset consists of 180-hour noise set which includes 150 classes 
and 65,000 noise clips, and over 500 hours clean speech, which includes audio clips from 2150 speakers. The clean speech dataset is derived from the public audiobooks
dataset called Librivox.
The noise clips were selected from Audioset and Freesound.
We randomly selected 24000 speakers 
with all noise clips to created 200-hour noisy train set,with singal-noise-ratio ranging from 0dB to 40dB.
Each selected speaker's audio clips are concatenated 
to 30 seconds,while mixing various noise clips. We estimated the proposed model 
with DNS-Challenge no-blind test dataset and blind test dataset.Both datasets consists of synthesis dataset 
and real-recordings.

\subsection{Training setup and baselines}
We trained the proposed model on both dataset using Adam optimizer annealed with warmup schedule, learning rate of 1e-3 and 64 mini-batchsize. The model was selected by early stopping when
the \(\frac{|\Delta Loss|}{Loss}\) was smaller than 0.05.
The STFT window length and hop size are modulated to 32 millisecond and 16 millisecond 
respectively. Hanning window and 512 FFT length are applied. We stacked the real and imaginary parts of FFT spectrogram
together as input.We compared the proposed model with several models,described as followed.
\begin{itemize}

\item \textbf{CRN}: A causal complex spectral mapping Convolution Recurrent Network\cite{CRN}, is evaluated on the first dataset as contrast.
This model has same structure as the proposed model CARN except attention mechanism.
It consists of six convolution layers encoder and  six symmetrical layers decoder with two lstm layers between thems.

\item \textbf{GCRN}: A causal complex spectral mapping Gate Convolution Recurrent Network\cite{GCRN}, is evaluated on the first dataset as constrast.This model has same structure as the CRN above,except gate convolution layers.

\item \textbf{DCCRN-E}: The model proposed by \cite{2020DCCRN}, which is based on the CRN architecture but 
complex convolution layers, has achieved the 1st place in the 2020 Deep Noise Suppression Challenge's
Real-Time Track. We compare this model under two datasets using officially opened  source code \footnote{ \url{https://github.com/huyanxin/DeepComplexCRN}}.

\item \textbf{CARN}: The proposed model,well described at section 2, is training and testing on both datasets,respectively.

\item \textbf{GCARN}: Same structure as the CARN, except gate convolution layers, is a contrast model.

\end{itemize}
\subsection{Evaluation:Objective Metrics and Result}

We evaluated wide-band Perceptual Evaluation of Speech Quality(PESQ)
-ITU-T P.862.2 \cite{PESQ},and the composite CSIG,CBAK,and COVL \cite{compositeMetrics} and STOI,on both test dataset.These metrics evaluated my model performance in some respects.
But these metrics require reference clean speech,and cannot work at real recordings.
The blind test dataset in DNS-Challenge 2020 consist of real recordings and synthetic noisy audios.
We evaluated our proposed models on this dataset with an another metric,called DNSMOS.

DNSMOS metric was proposed
by Chandan K A Reddy,etc.\cite{DNSMOS} recently.The DNSMOS metric is a non-intrusive objective speech quality metric for wide band scenario,and more reliable than other widely used objective metrics such as SDR and POLDA,
and does not require reference clean speech.So it can work on real recordings.

\begin{table}[!thpb]
	\caption{PESQ, CBAK, COVL,CSIG and STOI on Dataset 1}
	\label{tab:valentini_tab}
	\centering
	\footnotesize
	
	\setlength{\tabcolsep}{0.4mm}{
	\begin{tabular}{ccccccc}
		\toprule
		Method   & PESQ  & CBAK  & COVL & CSIG & STOI\\
		\hline
		Noisy    & 1.97 & 2.44 & 2.63  & 3.35 & 0.91\\
	  Wiener   & 2.22 & 2.68 & 2.67  & 3.23 & --\\
    SEGAN    & 2.16 & 2.94 & 2.80  & 3.48 & --\\
    U-Net    & 2.48 & 3.21	& 3.05	& 3.65 & --\\
    WaveNet  & --   & 3.24	& 2.98  & 3.62 & --\\
    CRN      & 2.61 & 3.26 &	3.17	& 3.78 & 0.94\\
    GCRN     & 2.51 & 3.24 & 3.09  &	3.71 &0.94 \\
    DCCRN-E	 & 2.73 &3.22  &3.22   &3.73	&0.941\\
    CARN     & 2.93 & $\mathbf{3.61}$	& $\mathbf{3.54}$	& 4.19 &0.95\\
    GCARN    &$\mathbf{2.99}$ & 3.46	&3.63	  & $\mathbf{4.27}$ &$\mathbf{0.96}$\\
		\hline
	\end{tabular}}
\end{table}

\begin{table}[!thpb]
  \caption{PESQ, CBAK, COVL,CSIG and STOI on Dataset 2
  }
  \label{tab:dns_tab}
  \centering
	\footnotesize
	\setlength{\tabcolsep}{0.4mm}{
	\begin{tabular}{ccccccc}
		\toprule
		Method   & PESQ  & CBAK  & COVL & CSIG & STOI\\
		\hline
    noisy	  &1.582	&2.533	&2.350	&3.186 & 0.92\\	
		RNNoise	&1.973	&3.463	&2.789	&2.692 &	--\\
    DNS-baseline	&1.81 &2.00	&2.23	&2.78 & --\\
    DCCRN-E    & 2.52 & 3.49 & 3.18  & 3.84  &0.94\\
poCoNet	&2.74	&3.04	&3.42	&4.08&--\\	
CARN  &2.91 &$\mathbf{3.67}$  &3.60	&$\mathbf{4.24}$	&0.96 \\
GCARN &$\mathbf{2.93}$  &3.65	&$\mathbf{3.61}$	&4.23	&$\mathbf{0.97}$\\
		\hline
	\end{tabular}}
\end{table}

In Table~\ref{tab:valentini_tab},
we listed the results over the first dataset 
of several methods ,such as Wiener,SEGAN\cite{2017SEGAN},U-Net\cite{2017Sing},
WaveNet\cite{2018WaveUnet},CRN,GCRN ,DCCRN-E and  our proposed models CARN and GCARN.
CRN and GCRN models were better than Wiener,SEGAN,U-Net and WaveNet,except CBAK scores,
and were not as good as DCCRN-E at PESQ.
We improved CRN and GCRN by replacing the directly connection with attention
between encoder layers and decoder layers,which we called CARN and GCARN models
respectively,achieving significant improvement in all these metrics.CARN improved 
CRN on PESQ,CBAK,COVL,CSIG with 12.2\%,10.7\%,11.7\%,10.8\% respectively,and 
GCARN improved GCRN on these metrics with 19.1\%,6.8\%,17.5\%,15.1\% respectively. CARN and GCARN were the best models among them.
When comparing 
CRN and CARN with GCRN and GCARN,the results demonstrated the gate convolution may be unecessary in this issue.

In Table~\ref{tab:dns_tab},we trained and evaluated CARN and GCARN models on DNS-Challenge dataset.We compared our proposed models with
RNNoise,the DNS-Challenge baseline\cite{reddy2020interspeech},DCCRN-E,and poCoNet \cite{Poconet} which 
took 1st place in the 2020 Deep Noise Suppression Challenge’s Non-Real-Time Track.
GCARN tied with CARN,and they achieved better scores in all these metrics,compared with these models.CARN outperformed DCCRN-E at PESQ,CBAK,COVL,CSIG with 15.5\%,5.2\%,13.2\%,10.4\% 
relatively,and outperformed poCoNet at these metrics with 6.2\%,20.7\%,5.3\%,3.7\% relatively.

In Table~\ref{tab:dns_dnsmos},we estimated our proposed models and DCCRN-E over the DNS-Challenge 2020 blind test dataset,
with the DNSMOS metric.CARN and 
GCARN can achieve significant imporvement speech quality compared with DCCRN-E.CARN scored the 
best in all non-reverb ,reverb and real-recordings.
\begin{table}[!thpb]
  \caption{DNSMOS over DNS-Challenge 2020 blind test dataset}
  \label{tab:dns_dnsmos}
  \centering
	\footnotesize
	\setlength{\tabcolsep}{0.4mm}{
	\begin{tabular}{cccccc}
		\toprule
		  Method & non-reverb  & reverb  & real-recordings & average \\
		\hline
    noisy	  &3.13	&2.64	&2.83 &2.95 \\	
    DCCRN-E &3.99	&3.16	&3.49	&3.53\\
		CARN	&$\mathbf{4.07}$	&$\mathbf{3.48}$	&$\mathbf{3.71}$	&$\mathbf{3.72}$   \\
    GCARN	&$\mathbf{4.07}$ &3.41	&3.69	&3.72 \\
		\hline
	\end{tabular}}
\end{table}

\section{Conclusion}
The experiments over the two datasets shows that attention mechanism can significantly improve
CRN architecture performance when compared with directly connection.
A reasonable explanation is
that attention mechanism filters some noise features which are connected from encoder layer
to decoder layer.Our proposed model CARN outperformed the place 1st model in both real time and non-real time track of the DNS Challenge 2020 at many metrics.



\bibliographystyle{IEEEtran}

\bibliography{mybib}


\end{document}